\documentclass[aps,prl,twocolumn,showpacs]{revtex4} 

\usepackage{graphicx,color}

\usepackage{amsthm}
\usepackage{amsmath}
\usepackage{amsfonts}
\usepackage{amssymb}

\usepackage[latin2]{inputenc}
\usepackage{mathptmx}
\usepackage{t1enc}

\DeclareMathOperator{\arth}{arth}
\renewcommand\Re{\operatorname{Re}}

\begin{document}
\title{Disentangling  dynamical phase transitions from equilibrium phase transitions}
\author{Szabolcs Vajna}
\affiliation{Department of Physics and BME-MTA Exotic  Quantum  Phases Research Group, Budapest University of Technology and
  Economics, 1521 Budapest, Hungary}
\author{Bal\'azs D\'ora}
\affiliation{Department of Physics and BME-MTA Exotic  Quantum  Phases Research Group, Budapest University of Technology and
  Economics, 1521 Budapest, Hungary}

\begin{abstract}
Dynamical phase transitions (DPT) occur after quenching some global parameters in quantum systems and are signalled  by the non-analytical time evolution of the dynamical free energy, which 
is calculated from the Loschmidt overlap  between the initial and time evolved states.
In a recent letter (M. Heyl et al., Phys. Rev. Lett. \textbf{110}, 135704 (2013)), it was suggested that DPTs are closely related to equilibrium phase transitions (EPT) for the transverse field Ising model. 
By studying a minimal model, the XY chain in transverse magnetic field, we show analytically that this connection does not hold generally. 
We present examples where DPT occurs without crossing any equilibrium critical lines by the quench, and a nontrivial example with no DPT but crossing a critical line by the quench. 
Albeit  the non-analyticities of the dynamical free energy on the real time axis do not indicate the presence or absence of an EPT, the structure of Fisher-lines for complex times reveal a qualitative difference. 
\end{abstract}
\date{\today}
\maketitle{}  
\bibliographystyle{apsrev}  

The interest about non-equilibrium dynamics grew immensely in the last few years \cite{PolkovnikovRMP,Barmettler2010,CalabreseTIM2012,Dziarmaga} 
thanks to experimental advances  made with ultracold atomic gases. The wide controllability of these systems allow experimentalists to prepare different 
kinds of non-equilibrium initial states and it is also possible to study the dynamics with time resolution unreachable in other physical systems \cite{Coldat1,Coldat2,Coldat3,Coldat4,Coldat5}. 
Some of the main questions are when and how thermalization, or  more generally equilibration, occurs and their connection to ergodicity and integrability. These were first posed by J. von 
Neumann already in 1929 \cite{Neumann1929}. 

The non-equilibrium time evolution can be characterized by many different ways, borrowing ideas from equilibrium statistical mechanics. 
{ The ultrashort time non-equilibrium dynamics, revealing the role of high-energy excitations, is also of interest as well as the stationary state reached after long time evolution. This latter} can be described by the diagonal ensemble,
 which is roughly the time averaged density matrix. 
The results of local measurements can be described by simpler ensembles, i.e. by the thermal Gibbs ensemble for non integrable (ergodic) systems \cite{BanulsTherm} 
and by the generalized Gibbs ensemble for integrable ones \cite{Rigol2007}. The Loschmidt overlap (LO), which is the main focus of this paper, 
is a nonlocal expression and is entirely determined by the diagonal ensemble, hence it characterizes the stationary state \cite{Fagotti2013_stac}. 
Analyzing the LO proved to be useful in studying quantum chaos, decoherence and quantum criticality \cite{Zanardi2006,Zanardi2007,Zanardi2011,Pollmann2010}. 
It is defined as the scalar product of the initial state and the time evolved state following a sudden global quench (SQ) as
\begin{equation} \label{eq:Zt}
G(t)=\left\langle \psi \right| e^{-i H t} \left|\psi \right\rangle,
\end{equation}
and can be regarded as the characteristic function of work performed on the system during the quench.
In a SQ the parameters of the Hamiltonian are changed suddenly from some initial to final values, and the system, prepared initially in the ground state $\left|\psi \right\rangle$ of the initial Hamiltonian, 
is assumed to be well separated from the environment and evolves unitarily with $H$. 

In a recent paper, Heyl et al.~\cite{dynQPT} pointed out a similarity between the time evolution of the LO overlap and the equilibrium phase transitions (EPTs). 
Close to phase transitions the free energy density is non analytical function of the temperature. A method proposed by Fisher \cite{Fisher1965} to analyze 
the zeros of the partition function in the \emph{complex temperature plane} gives a good understanding of these non-analyticities. 
In a finite system phase transitions cannot occur, and the Fisher-zeros are isolated and do not lie on the real axis. 
However, in the thermodynamic limit they coalesce into lines (or in general case areas \cite{Saarloos}) that can cross the real axis.
 These crossings are responsible for the breakdown of the analytic continuation of the free energy density as a function of temperature: 
knowing the free energy above the transition temperature does not give any informations about the free energy below. 

The LO in Eq.\eqref{eq:Zt} is formally similar to the canonical partition function with imaginary temperature. 
For a large system $G(t)$ scales exponentially with the system size, hence it is worthwhile to study the dynamical free energy {\cite{Fagotti2013_stac,PozsgayLecho}, which we define as} 
\begin{equation} 
\label{eq:dynfree} 
f(t)=-\lim_{N\rightarrow\infty}\frac 1N \ln G(t) \,.
\end{equation} 
Under certain circumstances this quantity shows non-analytical time-evolution. Due to the similarities with the EPT, the notion \emph{dynamical phase transitions} (DPTs) was introduced in Ref.~\cite{dynQPT}. 
It was found that in the transverse field Ising chain  the DPTs and EPTs are ultimately  related: 
the time evolution of $G(t)$ becomes non-analytic whenever the magnetic field is quenched through the (equilibrium) critical value. Similar observations were made for non-integrable  models~\cite{Karrasch2013} and for complex magnetic fields \cite{Hickey2013}.

The purpose of this paper is to show that this connection is not rigorous. To this aim we investigate the anisotropic XY chain in transverse magnetic field and show that generally DPTs 
can occur in quenches within the same phase, i.e. without crossing any equilibrium phase boundary. Note that a numerical evidence for this phenomenon was reported recently  
in Refs.\cite{Fagotti2013_stac,Sirker2013}. In addition, we also present a counter-example where the quench crosses an equilibrium critical point, but the LO remains analytic. 

The XY Hamiltonian with periodic boundary conditions reads as
\begin{equation} \label{eq:HamXY}
H(\gamma,h)=\sum_{j=1}^{N} \frac{1+\gamma}{2}\sigma_j^x\sigma_{j+1}^x+\frac{1-\gamma}{2}\sigma_j^y\sigma_{j+1}^y-h\sigma_j^z \,,
\end{equation} 
where $\gamma$ and $h$ are the anisotropy parameter and the homogeneous external magnetic field, respectively.  
This model can be mapped to free fermions with the use of Jordan-Wigner transformation as
\begin{align} \label{eq:HamXYfermi}
H(\gamma,h)=&\sum_{j=1}^{N-1}  c_j^{+}c_{j+1}+\gamma c_j^{+}c_{j+1}^{+}-h (c_j^{+}c_j-\frac{1}{2})+\textnormal{h.c.}\\
&-\mu (c_N^{+}c_1+\gamma c_N^{+}c_1^{+} +\textnormal{h.c}) \,, \nonumber
\end{align} 
where $c_j$ are fermionic operators and $\mu=e^{i \pi N_f}$, $N_f=\sum_{i=1}^{N}c_i^{+}c_i$. 
This Hamiltonian conserves the parity of the particle number and acts differently on the even and odd subspaces (sometimes referred to as Neveu-Schwarz or Ramond sectors). The Hamiltonian in the two subspaces are formally the same if we impose antiperiodic boundary condition for the even and periodic boundary  condition for the odd subspace. In wavenumber space these boundary conditions translate to different quantization of the wave numbers, $k=\frac{2\pi}{N} (n+\frac{1}{2})$ in the even and $k=\frac{2\pi}{N} n$ in the odd subspace.  In the fermionic language the ground state is unique in a given subspace, but when $|h|<1$ the ground states with even and odd parity become degenerate in the thermodinamic limit. 
These parity eigenstates are the symmetric or antisymmetric combinations of the fully polarized states, they do not possess magnetization in the coupling direction. 
We start our investigation with the parity eigenstates and we {discuss polarized ground states in the supplementary material}. 

The LO is calculated analytically in both of the even ($e$) and odd ($o$) subspaces as  
\begin{equation} \label{eq:ZtXY}
G_{s}(t)=e^{i \varphi_{s}(t)}\prod_{0<k<\pi}\left[ \cos(\epsilon_k t)+ i \cos(2\Theta_k)\sin(\epsilon_k t)\right] \,,
\end{equation}
where $\Theta_k=\theta_k^1-\theta_k^0$ is the difference between the Bogoliubov angles diagonalizing the pre- ($\alpha=0$) and post-quench ($\alpha=1$) Hamiltonians, $\epsilon_k\equiv\epsilon_k^1$ and for $s=o,e$, $\epsilon_k^{\alpha}=2\sqrt{(\cos(k)-h^{\alpha})^2+(\gamma^{\alpha}\sin(k))^2}$. 
The Bogoliubov angles are determined from $e^{i 2 \theta_k^{\alpha}}=2(\cos(k)-h^{\alpha}-i \gamma^{\alpha} \sin(k))/\epsilon^{\alpha}_k$, and
 the wave numbers are 
 quantized with respect to the parity of the initial state. The phase factor satisfies $\varphi_e(t)=0$ and $\varphi_o(t)=t(\pm \epsilon_0  \pm \epsilon_{\pi})/2 $, where the signs depend on the position of the initial and final Hamiltonian on the phase diagram \cite{suppmat}. 

We focus on the real part of the dynamical free energy, 
which is the same in thermodynamic limit for both sectors. 
The non analytical behaviour of the dynamical free energy is encoded in the zeros of the partition function $G(t)$ in the complex time plane~\cite{dynQPT}. Instead, following the practice in the literature we determine these zeros in the complex "temperature" plane, i.e. the zeros of  
$Z(z)=\left\langle \psi \right| e^{-z H} \left|\psi \right\rangle=G(-i t)$.
Specially in the XY model the Fisher zeros from $ Z(z)=0$ determine the dynamical free energy \emph{completely}~\cite{suppmat}. From Eq.~\eqref{eq:ZtXY}, 
 the Fisher zeros in the thermodynamic limit form lines indexed by an integer number $n$ as 
\begin{equation} \label{eq:FisherLine}
z_n(k)=\frac{i\pi}{\epsilon_k}(n+\frac{1}{2})-\frac{1}{\epsilon_k} \arth\left[\cos(2\Theta_k)\right]\,,
\end{equation} 
which agrees formally with Ref.\cite{dynQPT}, but in our case, the Bogoliubov angles depend on more variables, hence are  more general function of $k$. 
This increased freedom leads to interesting behaviour of the Fisher lines. The main quantity that determines the dynamical free energy is $\cos(2\Theta_k)$, which 
can be expressed analytically with the parameters of the initial and final Hamiltonian.  
Furthermore, $\cos(2\Theta_{k})=1-2n_k$, where $n_k$ is the expectation value of the quasiparticle occupation number in the post-quench Hamiltonian and is conserved under the time evolution. 
A Fisher line crosses the imaginary axis whenever $n_k=1/2$, which can be interpreted as modes with infinite effective temperature. 
These crossings are responsible for the non analytic time evolution of the dynamical free energy.
   
Due to the parity of the cosine function it is evident that if a Fisher line crosses the imaginary axis for a quench $(h_0,\gamma_0)\rightarrow(h_1,\gamma_1)$ it implies a crossing in the 
reversed protocol $(h_1,\gamma_1)\rightarrow(h_0,\gamma_0)$ as well. We call this as the symmetric property of DPT. This seems to be plausible in quenches crossing 
critical points, but it is less trivial for quenches within the same phase.

\begin{figure}[h!]
\centering
\includegraphics[width=8.6cm]{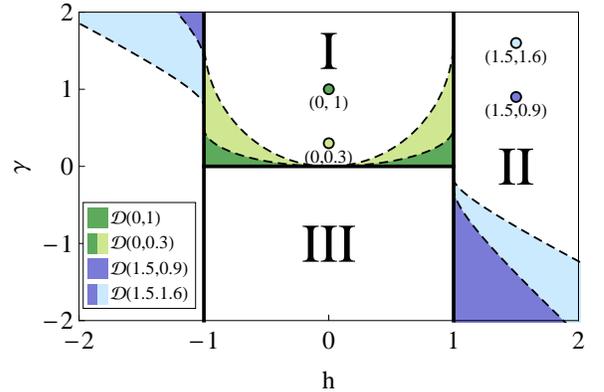}
\caption{The phase diagram of the XY model in magnetic field. The three studied phases (I,II,III) are marked on the plot.
These gapless phases are separated by critical lines that form an "H" letter-like shape.  DPTs can
occur in quenches within the same phase. The domains $\mathcal{D}(h_0,\gamma_0)$ of the final parameters where DPTs 
appear are shown for four given initial conditions $(h_0,\gamma_0)$. Except from the region $h_1<-1$ the domains are determined from Eq.\eqref{eq:Dh0g0}.
Note that when quenching from $II$ to $h_1<-1$, non-analyticities only show up in the top-left corner of the phase diagram and remain absent otherwise, in spite of crossing several critical lines. }
\label{fig:phasediag}
\end{figure}

The phase diagram of the XY chain is drawn on Fig.\ref{fig:phasediag}. The excitation spectrum is gapless when $h= \pm 1$ or when $\gamma=0$, $|h|<1$. The Fisher-lines, and hence the LO 
 show different behaviour for quenching between different regions in the phase diagram. The exact values of the initial and final parameters $h_0$, $\gamma_0$, $h_1$, and $\gamma_1$ 
in given phases do not modify qualitatively the behaviour of the LO as a sign of some kind of universality.   
We consider 4 types of quenches, 3 of them can be realized with quenching one parameter only, while in third example one needs to quench both the magnetic field and the anisotropy parameter. 

\begin{figure}[h!]
\centering
\includegraphics[width=2.6cm]{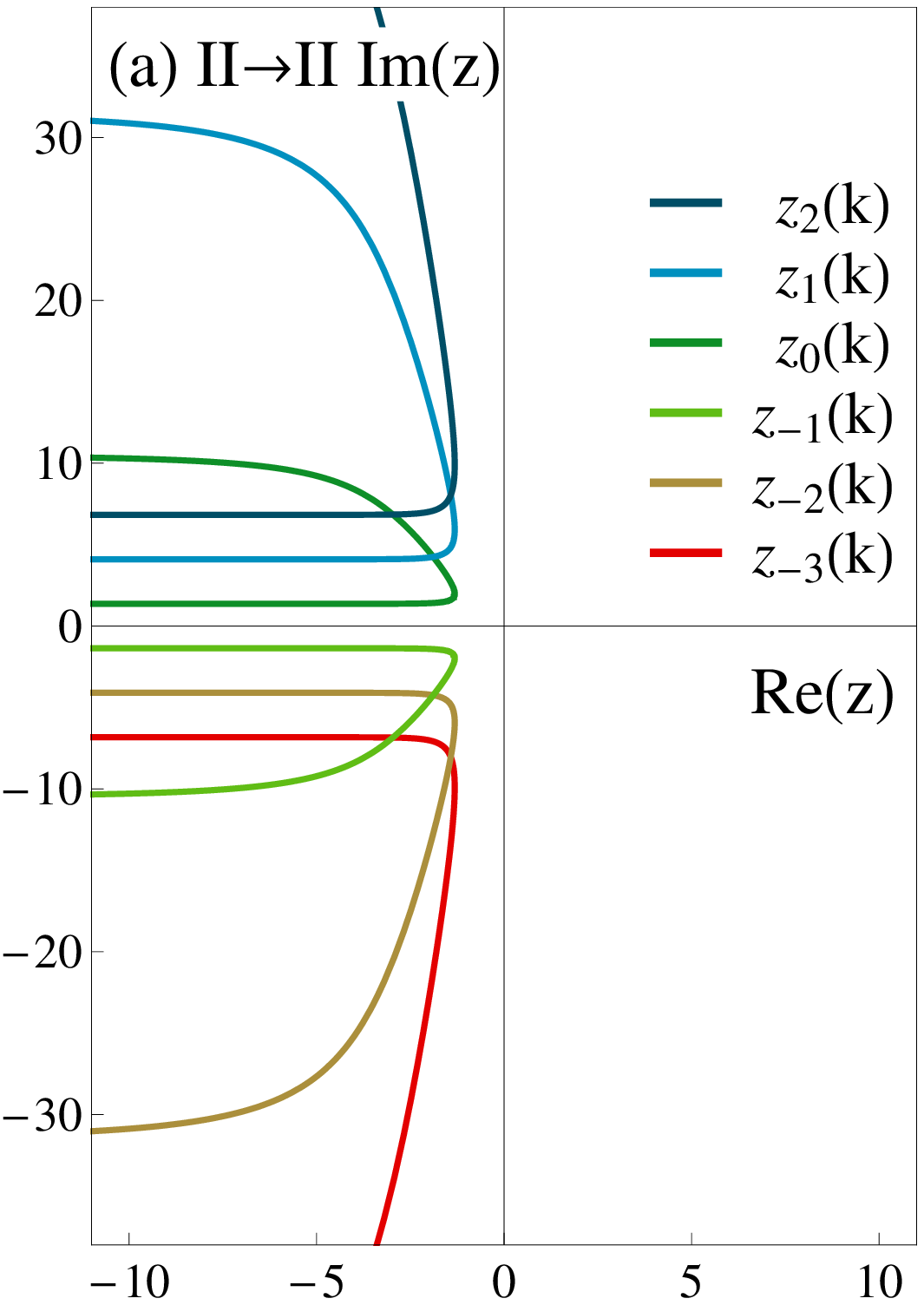}
\includegraphics[width=2.6cm]{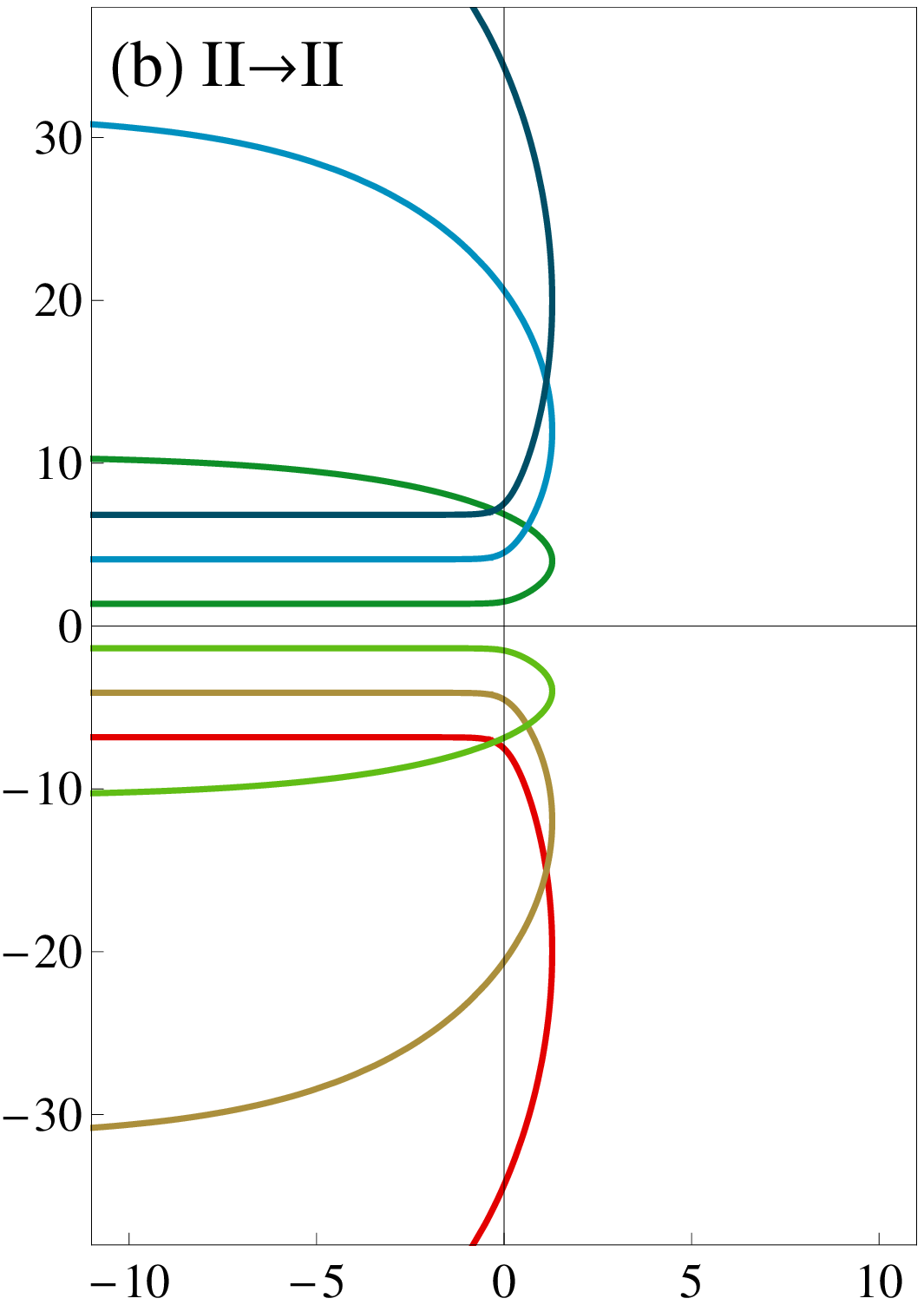}
\includegraphics[width=2.6cm]{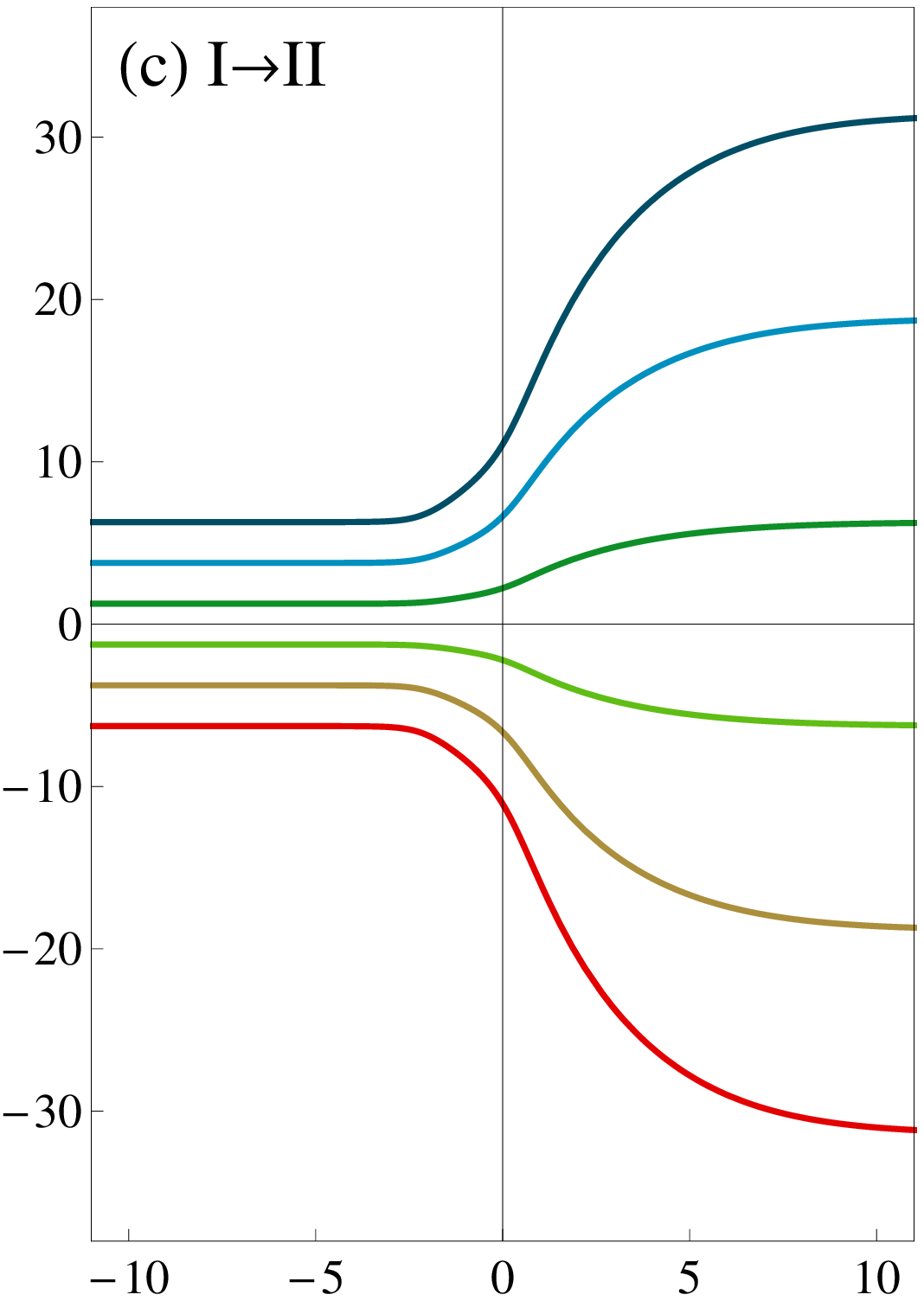}
\includegraphics[width=2.6cm]{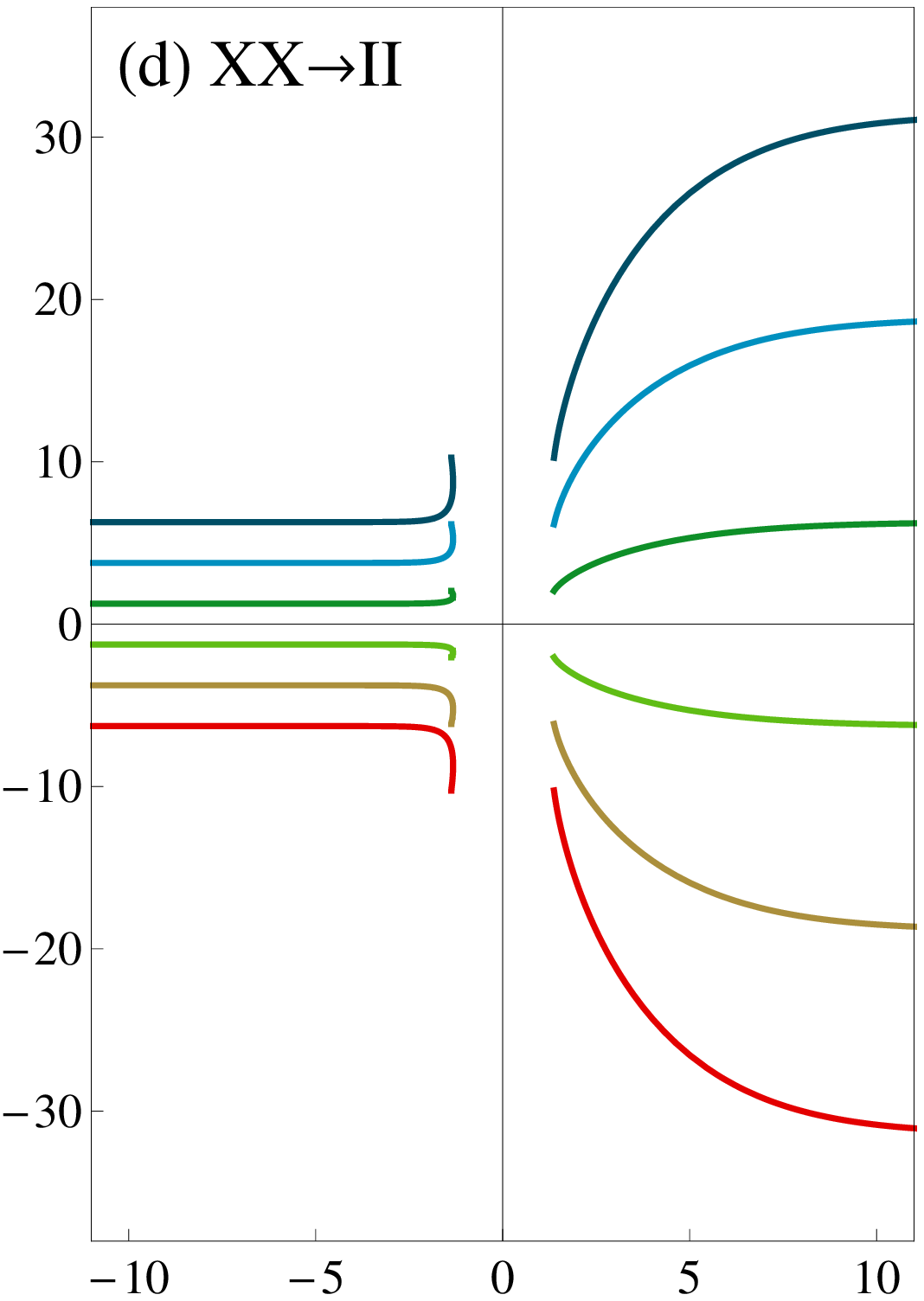}
\includegraphics[width=2.6cm]{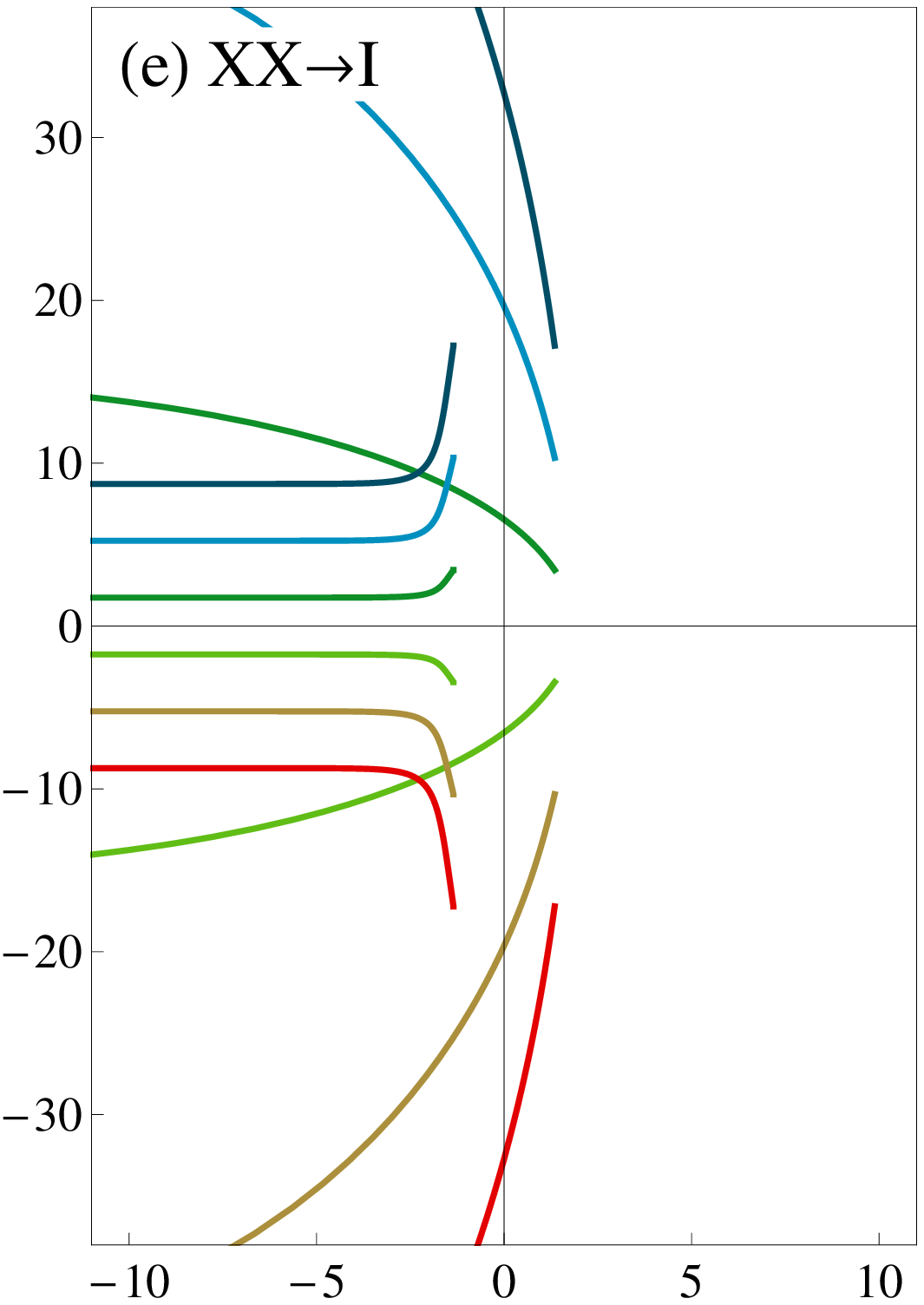}
\includegraphics[width=2.6cm]{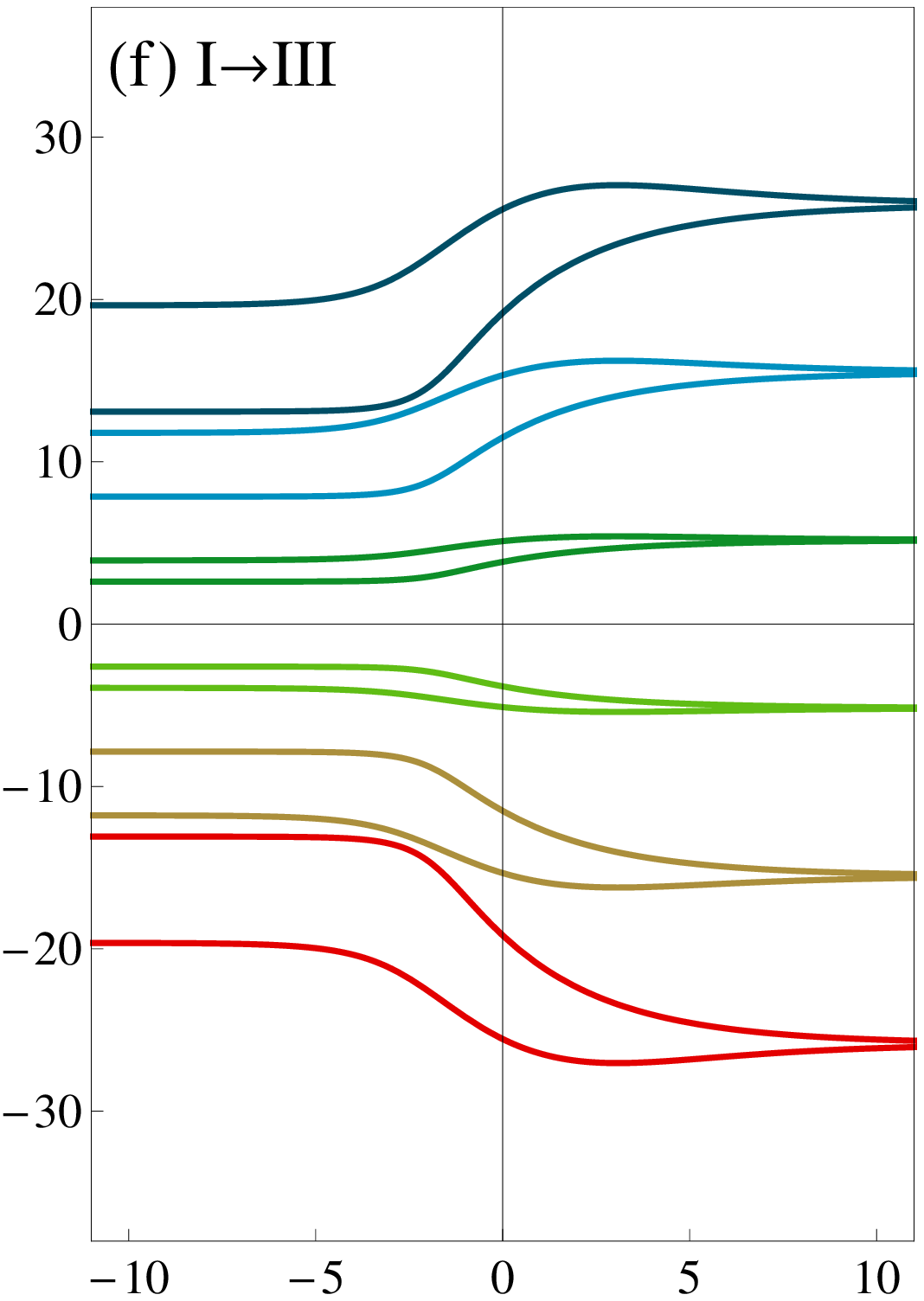}
\caption{The flow of Fisher-lines $z_{n}(k)$ ($n=(-3,\dots,2)$) for various types of quenches discussed in the main text.}
\label{fig:fisherline}
\end{figure}

\emph{DPT without EPT: Quenches not crossing critical points}. 
We start our discussion with quenches inside phase $II$, where $h_{0,1}>1$ and we assume that $\gamma_0>0$ without loss of generality. In this setup no critical lines are crossed by the parameters of the 
Hamiltonian during the quench, but DPTs can occur. 
Generally one can  show that the $k\rightarrow 0,\pi$ tails of the Fisher lines lie in the left half plane: $\lim_{k\rightarrow0} \Re\{z_n(k)\}= -\infty$ and $\lim_{k\rightarrow\pi} \Re\{z_n(k)\}= -\infty$. 
For small quenches the whole lines lie in the left half plane (Fig.\ref{fig:fisherline}a), hence the time evolution of the dynamical free energy is analytical. However, the turning point of the Fisher lines 
can move to the right half plane (Fig.\ref{fig:fisherline}b). In this case each Fisher line crosses the time axis twice at wave numbers $k_1^{*}$ and $k_2^{*}$. 
The non-analytical times are given by  $t_i^{*}=\frac{\pi}{\epsilon_{k_i^{*}}}(n+\frac{1}{2})$, $i=1,2$. This occurs if the anisotropy parameter is quenched 
to a sufficiently negative value at fixed magnetic field. No matter how $\gamma$ is quenched, an equilibrium critical point is never approached, but DPT shows up.

{More generally for each point $(h_0,\gamma_0)$ in phase $II$ , the domain $\mathcal{D}(h_0,\gamma_0)\subset II$ of $(h_1,\gamma_1)$, where DPT occurs is given by}
\begin{small}
\begin{equation} \label{eq:Dh0g0}
\mathcal{D}(h_0,\gamma_0)=\{(h_1,\gamma_1)| 2\gamma_0\gamma_1<1-h_0 h_1-\sqrt{(h_0^2-1)(h_1^2-1)}\}
\end{equation}
\end{small}
within phase $II$. 
The boundary of these regions is a second order curve (a cone-section).  A few examples for these domains are shown on Fig.\ref{fig:phasediag}.

Similar phenomenon can be observed in quenches inside phase $I$. The Fisher lines start and end in the left half of the complex plane, but some parts of the lines can move to 
the right half plane. Given $(h_0,\gamma_0)$ in phase $I$ the domain of final parameters where the non analyticities occur is given by Eq.\eqref{eq:Dh0g0} within phase $I$. 
For example starting from the Ising model ($\gamma_0=1$, $h_0=0$) one needs to quench the magnetic field and the anisotropy parameter as well to see the non-analytic behaviour (see Fig.\ref{fig:phasediag}). 
However, considering smaller anisotropy 
the DPT can  appear by quenching solely the magnetic field when $\gamma_0<\sqrt{1+|h_0|}/\sqrt{2}$ is satisfied for the initial Hamiltonian. 

\emph{DPT together with EPT: Quench between phases I and II.} In this setup the quenched parameters cross at least one critical point, and the time evolution of the dynamical free energy is always non analytical. 
The asymptotic behaviour of the Bogoliubov angles guarantee that the Fisher lines cross the imaginary axis, that is,  $\lim_{k\rightarrow0} \Re\{z_n(k)\}= \infty$ and $\lim_{k\rightarrow\pi} \Re\{z_n(k)\}= -\infty$ (
Fig.\ref{fig:fisherline}c). Because of the symmetries of the XY model quenches between phase II and III behave in the same way. 

\emph{EPT without DPT: Quench from phase II to the critical XX line ($\gamma=0$, $|h|<1$).} 
In quenches $II\rightarrow I,III$  DPTs showed up everywhere except for quenches from phase $II$ to the boundary of $I$ and $III$.  
Though the asymptotic behaviour of the Bogoliubov angles are similar to the $I\rightarrow II$ case, there is an interesting difference as well: there are no Fisher zeros in the vicinity of the imaginary axis. 
The function $\cos(2\Theta_k)$ is not continuous at $\tilde{k}=\arccos(\frac{h_1 \gamma_0-h_0 \gamma_1}{\gamma_0-\gamma_1})$, therefore  $\lim_{\epsilon\rightarrow 0^+}\cos(2\Theta_{\tilde{k}\mp\epsilon}) \lessgtr 0 $. 
Hence the Fisher lines split into two sections that do not cross the imaginary axis (Fig.\ref{fig:fisherline}d). 

By considering the XX line as the $\gamma_1\rightarrow 0$ limit, then as we approach the XX line the slope of $\cos(2\Theta_k)$ diverges at $\tilde{k}$, hence the density of Fisher zeros vanish near the imaginary axis. 
As opposed to previous examples, when the initial and final Hamiltonians lied in the gapped phase, 
quenching to the XX line is a special case because the final parameters are on a critical line. Nevertheless, it is still surprising that for quenches $II\rightarrow I,III$ DPTs occur everywhere except for the boundary of these regions. 

However, non analytical behaviour in the dynamical free energy can be observed in quenches to the critical lines as well. One example is a quench from I or III to the XX line: 
$(\gamma_0\neq 0,|h_0|<1)\rightarrow (\gamma_1=0,|h_1|<1)$ with $h_1\neq h_0$. In this case, one would think  naively the Fisher lines would cross the imaginary axis 
twice similarly to quenches $I\rightarrow I$ {and $I\rightarrow III$}, but one of the crossings does not manifest itself (Fig.\ref{fig:fisherline}e) in a similar manner as it was discussed in the previous paragraph.  
The other example, which we only mention here, is a quench crossing a critical line \cite{qcross_expl}: starting from $III$ to the $h=1$ critical boundary of $I$.  

\begin{figure}[htb!]
\centering
\includegraphics[width=4.28cm]{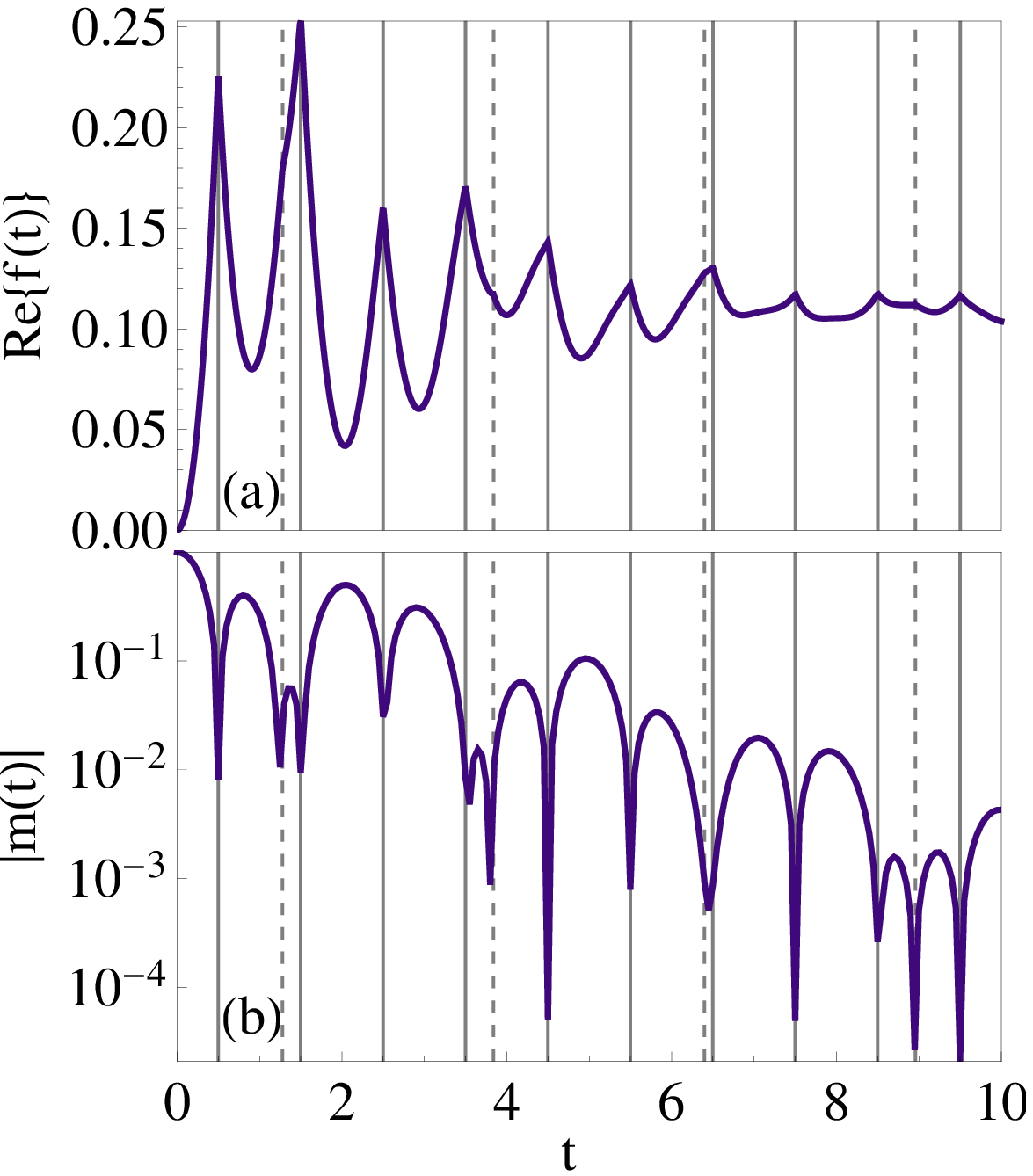}
\includegraphics[width=4.28cm]{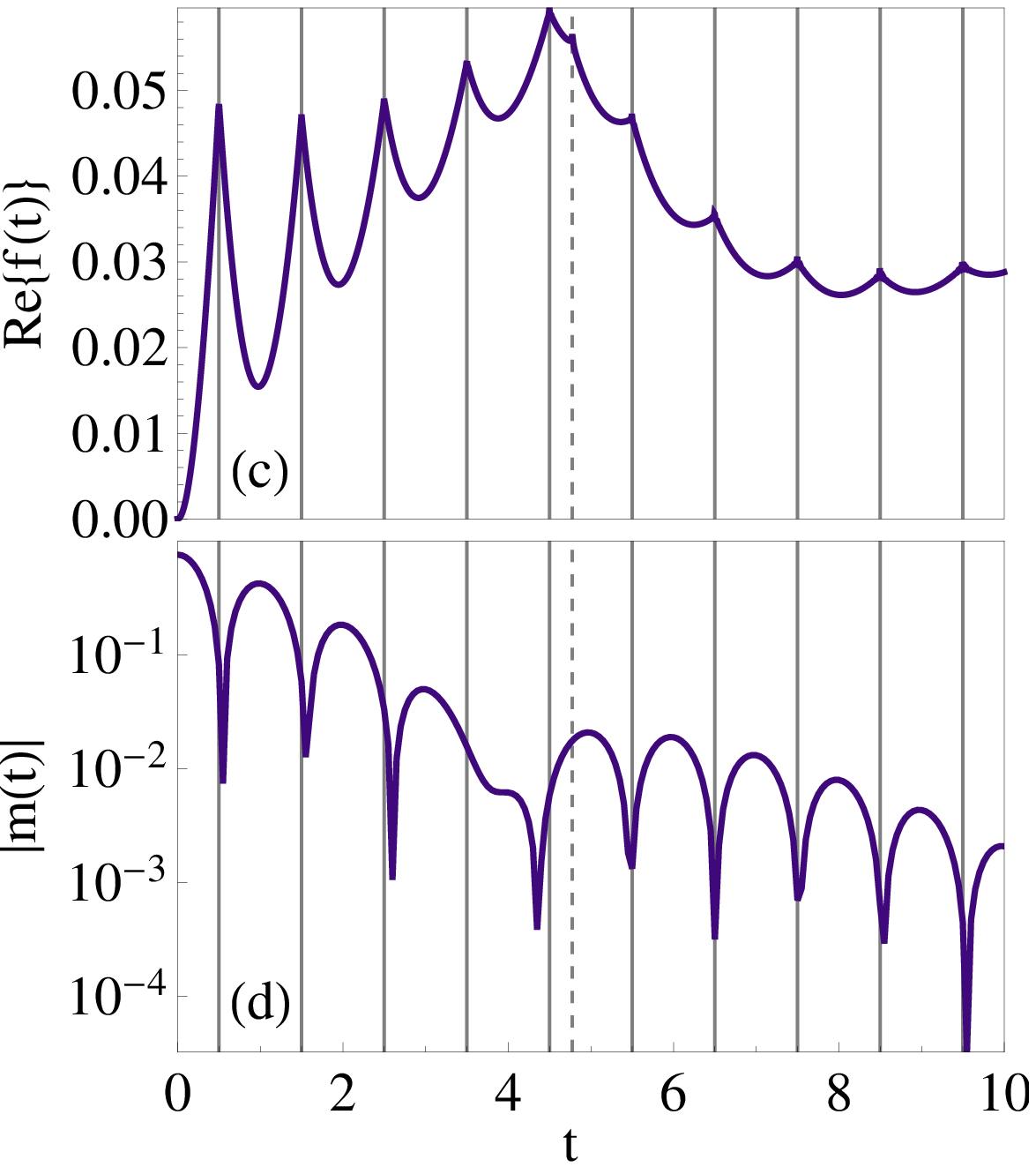}
\caption{The dynamical free energy is non analytical at Fisher times $t_{i,n}=t_{i}^{*}(n+1/2)$ $i=1,2$ (solid and dashed lines respectively). The time unit was chosen to be $t_{1}^{*}$. The longitudinal magnetization also shows two timescales, in case b) the zeros of the magnetization approximately lie at the Fisher times, in d) the relation between them is more involved. Quench parameters for a) and b) are $(h_0=0,\gamma_0=1)$ $\rightarrow$ $(h_1=0.6,\gamma_0=-1)$ and for c) and d) are $(h_0=0,\gamma_0=0.1)$ $\rightarrow$ $(h_1=0.6,\gamma_0=0.1)$}
\label{fig:dynfree_magn}
\end{figure}

\emph{Quench from phase I to III}. In this case the anisotropy parameter is quenched from positive to negative values in low 
magnetic field ($-1<h_{0,1}<1$). The system goes through an anisotropy transition at $\gamma=0$. At $\gamma>0$ the ground state 
polarization is in the $x$, while at $\gamma<0$ it is in the $y$ direction. For these quenches $\lim_{k\rightarrow0,\pi} \Re\{z_n(k)\}= -\infty$ 
meaning that the Fisher lines start and end at the left half plane. However, there is wavenumber $0<\tilde{k}<\pi$ defined by 
$\cos(\tilde{k})=\frac{h_1 \gamma_0-h_0 \gamma_1}{\gamma_0-\gamma_1}$, for which $\cos(2\Theta_{\tilde{k}})=-1$. This means that 
while $k$ goes through the interval $(0,\pi)$ the Fisher lines come from $\Re\{z\}=-\infty$, reach $\Re\{z\}=\infty$ at $\tilde{k}$ 
and finally go back to $\Re\{z\}=-\infty$ again (Fig.\ref{fig:fisherline}f). Hence all the Fisher lines cross the imaginary axis twice 
giving rise to two emergent timescales in the dynamical free energy (Fig.\ref{fig:dynfree_magn}a). This is qualitative difference between 
the quenches I to II and I to III.

For EPTs, the non-analyticity of the free energy is also imprinted in the non-analytic behaviour of other physical quantities, e.g. the order parameter or its susceptibility.
A similar phenomenon is expected to occur for the DPTs as well~\cite{dynQPT}. 
For the XY model, the equilibrium order parameter is the magnetization in the XY plane. Therefore, we determined its absolute value for the non-equilibrium situation by numerical evaluation of Pfaffians \cite{Barouch1971}.
Whenever the Fisher line crosses the imaginary axis once, only a single emergent non-equilibrium timescale appears from the dynamical free energy, which
matches exactly that of the magnetization.
However, for quenches $I\rightarrow I$ and $I\rightarrow III$ each Fisher line crosses the imaginary axis twice which implies two non equilibrium timescales. 
Only these two timescales and their higher harmonics (in Fig.~\ref{fig:dynfree_magn}d) appear in the dynamics of magnetization, though generally we were not able to express analytically
the zeros of the magnetization by the non analytic timescales. 
However, in the $I\rightarrow III$ quench protocol when $\gamma_0$ and $\gamma_1$ 
are not too close to the $\gamma=0$ critical line, the magnetization takes zero values in the vicinity of the Fisher times (Fig. \ref{fig:dynfree_magn}a,b). 

Until now we considered quenches starting from even or odd parity eigenstates. It is an important question whether the non analytic behaviour is present in quenches starting from polarized states or not. For quenches through the critical point in the transverse field Ising model it has been shown that DPTs can be observed, but the non-analyticities are not at the Fisher-times of the parity subspaces \cite{dynQPT,Karrasch2013}. We found similar behaviour in the XY  model \cite{suppmat}. 

Though we calculated the LO and the dynamical free energy directly from the time evolution of the initial wave function, they describe the stationary state after the quench \cite{Fagotti2013_stac}. That is, as the time evolution operator is diagonal in the eigenbasis of $H$, 
$G(t)$ depends only on the diagonal elements of the density matrix:  $G(t)=Tr \{ \rho_{DE} e^{-i H t} \}$, where $\rho_{DE}$ is the diagonal ensemble. The diagonal density matrix depends on the fermion occupation numbers $n_k$ and it can be expressed explicitly \cite{suppmat}
\begin{align} \label{eq:rho_DE}
\rho_{DE}&=\prod_{0<k<\pi}(n_k n_{-k}+\cos^2(\Theta_k)(1-n_k-n_{-k}))\\
&=\prod_{0<k<\pi} \cos^2(\Theta_k) \delta_{n_{k},0}\delta_{n_{-k},0}+\sin^2(\Theta_k) \delta_{n_{k},1}\delta_{n_{-k},1} \label{eq:rho_DE2}
\end{align}
From the latter form it is straightforward to reproduce Eq. ~\eqref{eq:ZtXY}. 
The correlation between wave numbers $k$ and $-k$ come from the BCS superconductor-like initial state. The LO - up to a trivial phase factor - is the characteristic function of work done on the system \cite{rev_qfluct}, 
hence it depends on all moments of the energy. As it is a non-local quantity the generalized Gibbs ensemble $\rho_{GGE}\sim e^{\sum\lambda_k n_k}$, where $\lambda_k$ fixes the expectation value of $n_k$, would not give 
the proper result for the LO, because it does not describe well the correlations between $n_k$ and $n_{-k}$. With the diagonal ensemble in Eq.~\eqref{eq:rho_DE},
 we took into account the correlations among the modes hence it can be applied to calculate any  moment of the energy.

\emph{Conclusion.} We analyzed the dynamical free energy for quenches in the XY model in magnetic field. The singular behaviour of the dynamical free energy is determined solely 
by the Bogoliubov angles through the quasiparticle occupation numbers and it is not sensitive to the spectra of the initial or final Hamiltonians. The appearance of DPTs 
are connected to the existence of modes with $1/2$ occupancy probability. 
In this particular system we explicitly demonstrated the existence of DPTs without an EPT as well as the absence of DPTs in the presence of EPTs. 
Though the dynamical free energy does not distinguish between DPTs with or without EPTs, the Fisher lines do. 
If the quench crosses a critical line, the Fisher lines sweep through the whole real axis. However, for quenches inside a given phase, the Fisher lines reach either $\infty$ or $-\infty$. 

\begin{acknowledgments}

This research has been  supported by the Hungarian Scientific  Research Funds Nos. K101244, K105149, K106, CNK80991, by the ERC Grant Nr. ERC-259374-Sylo and 
by the Bolyai Program of the Hungarian Academy of Sciences.
\end{acknowledgments}

\appendix
\section{Supplementary material for "Distinct occurance of dynamical phase transitions from equilibrium phase transitions" (Sz. Vajna, B. D\'{o}ra)}
\setcounter{equation}{0}
\renewcommand{\theequation}{S\arabic{equation}}

\setcounter{figure}{0}
\renewcommand{\thefigure}{S\arabic{figure}}

\section*{Initial state of the parity sectors}
\begin{figure}[htb!]
\centering
\includegraphics[width=8.8cm]{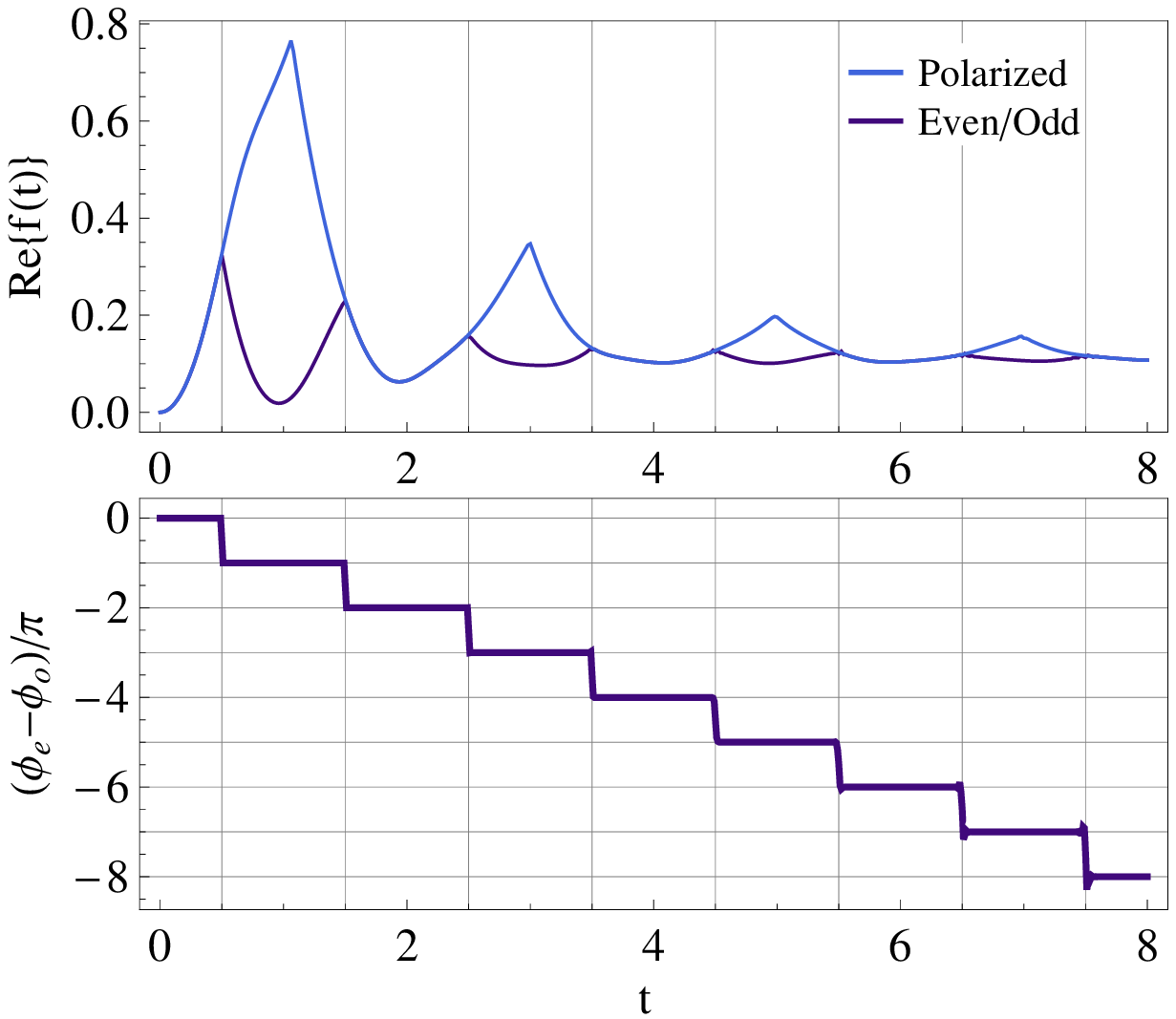}
\caption{Top: dynamical free energy from parity eigenstates and polarized initial states. Bottom: the phase difference of the Loschmidt overlaps in the even and the odd sectors.}
\label{fig:pol}
\end{figure}

In this section we give more details on the ground states of the XY model in the even and odd parity sector and we give explicitly the signs in the phase factor $\varphi_o$ in Eq.(5) of the main text. The explicit form of the initial wave function allow one to construct the density matrix and from Eq.\eqref{eq:rho_op} the diagonal density matrix (Eq.(9) of the main text) immediately follows.
As we discussed in the main text the XY Hamiltonian in the fermionic representation acts differently on the even and odd quasiparticle sectors. In the even sector we imposed anti-periodic boundary conditions on the fermions, hence $k\in \{ \pm\frac{\pi}{N},\pm\frac{3\pi}{N},\dots,\pm\frac{N-1}{N}\pi \}$, while in the odd sector $k\in \{0, \pm\frac{2\pi}{N},\pm\frac{4\pi}{N},\dots,\pi \}$. 
In the odd sector one needs to separate wave numbers $k=0$ and $k=\pi$, because for these $-k$ is identical to $k$. Following a quench the initial state can be expressed in terms of the fermions $f_k$ that diagonalize the new Hamiltonian: $H=\sum_k \epsilon_k (f_k^{+} f_k-1/2)$ for both sectors. 
In the even subspace
\begin{align} \label{eq:GSe}
\left|GS_e\right\rangle=\prod_{0<k<\pi}\cos{\Theta_k}-\sin{\Theta_k} f_k^{+}f_{-k}^{+} \left|0\right\rangle_f^e \,,
\end{align}
where the product contains all the wave numbers and $\left|0\right\rangle_f^e$ denotes the vacuum of the $f_k$ quasiparticles of the even subspace. In the odd sector the vacuum of the $f_k$ quasiparticles lies in the odd subspace if $|h|<1$ and in the even subspace if $|h|>1$.
\begin{subequations} 
\begin{align} 
&\left|GS_o\right\rangle=& \hspace{-2em} \prod \cos(\Theta_k)-\sin{\Theta_k} f_k^{+}f_{-k}^{+} \left|0\right\rangle_f^o \\
&\left|GS_o\right\rangle=& \hspace{-2em} f_{\pi}^{+} \prod \cos(\Theta_k)-\sin{\Theta_k} f_k^{+}f_{-k}^{+} \left|0\right\rangle_f^o  \\
&\left|GS_o\right\rangle=& \hspace{-2em} f_0^{+} \prod \cos(\Theta_k)-\sin{\Theta_k} f_k^{+}f_{-k}^{+} \left|0\right\rangle_f^o \\
&\left|GS_o\right\rangle=& \hspace{-2em} f_0^{+} f_{\pi}^{+}  \prod \cos(\Theta_k)-\sin{\Theta_k} f_k^{+}f_{-k}^{+}\left|0\right\rangle_f^o   
\end{align}
\end{subequations}
where the product contains all the possible wave numbers except $0$ and $\pi$, and (a) corresponds to quenches inside phase I or phase II, (b) to quenches from I or III to the phase $h<-1$, (c) to quenches from I or III to II, and (d) to quenches from II to phase $h<-1$. 
With the knowledge of the initial states the Loschmidt overlap in Eq.(5) of the main text can be reproduced easily. The phase factors for the cases discussed above are 
\begin{subequations}
\begin{align} 
\varphi_o(t)&=t(- \epsilon_0  - \epsilon_{\pi})/2 \\
\varphi_o(t)&=t(- \epsilon_0  + \epsilon_{\pi})/2 \\
\varphi_o(t)&=t(+ \epsilon_0  - \epsilon_{\pi})/2 \\
\varphi_o(t)&=t(+ \epsilon_0  + \epsilon_{\pi})/2  \,.
\end{align}
\end{subequations}
This phase factor plays an important role in the Loschmidt overlap (LO) of the polarized states, without this term one would not get the step function for the phase difference between the LO of the even and odd subspace (Fig.~\ref{fig:pol}).

\section*{Loschmidt overlap for polarized states}
The polarized states are superpositions of the even and odd ground states: $\left|GS_{pol,\pm}\right\rangle = \frac{1}{\sqrt{2}}(\left|GS_{e}\right\rangle \pm \left|GS_{o}\right\rangle)$. The Hamiltonian conserves the parity, hence the LO of the polarized state is simply the average of overlap calculated in the even and odd ground states:
\begin{equation}
G_{pol}(t)=e^{i \phi_{e}(t)}|G_{e}(t)|+e^{i \phi_{o}(t)}|G_{o}(t)|
\end{equation}
where we decomposed the complex overlap to the absolute value and a phase factor. 
The overlap can be evaluated numerically, and we found that the dynamical phase transitions occur in the polarized case as well. 
We observed that the phase difference $\phi_{e}(t)-\phi_{o}(t)$ is a piecewise constant function in the thermodynamic limit. At $t=0$ the phase difference is zero, and it is changed by $\pi$ at the Fisher points (see Fig.1). This means that depending on the time $|G_{pol}(t)|=\left| |G_{e}(t)| \pm |G_{o}(t)| \right|$. Though real parts of the dynamical free energies $f_e$ and $f_o$ are equal in the thermodynamic limit, this does not imply that $|G_{pol}(t)|$ would be identically zero at certain intervals, because of the division by the system size in the definition of the dynamical free energy. In other words  $G_{e/o}$ cannot be approximated by $e^{-N f_{e/o}(t)}$ when we subtract the two terms. The non-analytic behaviour of the dynamical free energy for polarized initial state is encoded in the small difference between $|G_{e}|$ and $|G_{o}|$.

\section*{Density matrix}
From Eq.\eqref{eq:GSe} the initial and the time evolved density matrix can be constructed.
\begin{small}
\begin{align} \label{eq:rho_op}
\rho(t)=&\prod_{0<k<\pi}  \left\{ n_k n_{-k}-\cos^2{\Theta_k} (n_k+n_{-k}) +\cos^2{\Theta_k} \right. \\
& \left. - e^{i \epsilon_k t}\cos{\Theta_k} \sin{\Theta_k}  f_{-k} f_k - e^{-i \epsilon_k t}\cos{\Theta_k} \sin{\Theta_k} f_{k}^{+} f_{-k}^{+} \right\}
\end{align}
\end{small}
The density matrix is the tensor product of the reduced density matrices of $(-k,k)$ subspace. In the basis $\left|n_k n_{-k}\right\rangle$, that is $\left|00\right\rangle$, $\left|01\right\rangle$,$\left|10\right\rangle$ and  $\left|11\right\rangle$ the density matrix looks like
\begin{small}
\begin{align}
\rho(t)=\prod_{0<k<\pi}\otimes 
\begin{pmatrix} 
\cos^2{\Theta_k} & 0 & 0 & - e^{i \epsilon_k t}\cos{\Theta_k} \sin{\Theta_k} \\
0 & 0 & 0& 0\\
0 & 0 & 0& 0\\
- e^{-i \epsilon_k t}\cos{\Theta_k} \sin{\Theta_k} & 0 & 0 & \sin^2{\Theta_k}
\end{pmatrix}
\end{align}
\end{small}
In the time averaged density matrix $\bar{\rho}=\lim_{T\rightarrow\infty}\frac{1}{T}\int_0^T \rho(t) \mathrm{d}t$ the diagonal elements will remain the same, and almost all off-diagonal elements vanish except for degenerate wave numbers, which possess the same energy. For example when $h^1=0$, $\epsilon_k=\epsilon_{\pi-k}$ hence the process $f_{\pi-k}^{+} f_{k-\pi}^{+}f_{k} f_{-k}$ is non vanishing in the stationary state. As the time evolution is diagonal in the quasiparticles $f_k$, the non-diagonal elements of $\bar{\rho}$ does not influence the Loschmidt overlap, but for the stationary expectation value of a general physical quantity one should take into account these terms.

\section*{Loschmidt overlap from the Fisher lines}
In \cite{dynQPTa} it was shown that the Fisher zeros solely determine the non-analytic part of dynamical free energy
\begin{align} \label{eq:fz_fisher}
f(t)=-\lim_{N\rightarrow \infty}\frac{1}{N} \left[h(t)+\sum_{n,k} \ln\left(1-\frac{i t}{z_{n}(k)}\right)\right]
\end{align}
where $t$ denotes time, but can be extended to complex values, $z_{n}(k)$ are the Fisher zeros determined in Eq.(6) of the main text and $h(t)$ is an entire function depending on the model under consideration. In the XY model $h(t)=i \varphi_{e/o}(t)$ in the even/odd sector. To see this we use the infinite product representation of the sine function $\sin(x)=x\prod_{n=1}^{\infty}\left(1-\frac{x^2}{\pi^2 n^2}\right)$ \cite{AbrStegun} to prove that
\begin{align}
\prod_{n=-\infty}^{\infty}\left(1-\frac{2x}{\pi(2n+1)+2a}\right)=\cos(x)+ \sin(x)\tan(a)
\end{align}
Now we substitute $z_n(k)$ and use this identity to the singular part of $f(t)$ in Eq.\eqref{eq:fz_fisher}
\begin{align}
\prod_{n=-\infty}^{\infty}\left(1-\frac{i t}{z_{n}(k)}\right)=\cos(\epsilon_k t)+i \cos(2\Theta_k) \sin(\epsilon_k t) \,.
\end{align}
This is the general $k$ term in the formula for the Loschmidt overlap in Eq.(5) of the main text, hence the entire function $h(t)=i \varphi_{e/o}(t)$. Because it is purely imaginary for real times, the Fisher zeros entirely determine the (important) real part 
of the dynamical free energy.


\begin{thebibliography}{99}
\bibitem{qcross_expl} By this we mean that any curves connecting the initial and final parameters cross at least one critical lines.
\end{thebibliography}
\end{document}